\documentclass[10pt,a4paper,twocolumn]{article}
\usepackage{graphicx}

\begin{document}

\title{The particle invariance in particle physics}
\author{H. Y. Cui\\
Department of Applied Physics\\
Beijing University of Aeronautics and Astronautics\\
Beijing, 100083, China}
\date{( Oct., 15, 2001 )}

\maketitle

\begin{abstract}
{Since the particles such as molecules, atoms and nuclei are
composite particles, it is important to recognize that physics must be
invariant for the composite particles and their constituent particles, this
requirement is called particle invariance in this paper. But difficulties
arise immediately because for fermion we use Dirac equation, for meson we
use Klein-Gordon equation and for classical particle we use Newtonian
mechanics, while the connections between these equations are quite indirect.
Thus if the particle invariance is held in physics, i.e., only one physical
formalism exists for any particle, we can expect to find out the differences
between these equations by employing the particle invariance. As the
results, several new relationships between them are found, the most
important result is that the obstacles that cluttered the path from
classical mechanics to quantum mechanics are found, it becomes possible to
derive the quantum wave equations from relativistic mechanics after the
obstacles are removed. An improved model is proposed to gain a better
understanding on elementary particle interactions. This approach offers
enormous advantages, not only for giving the first physically reasonable
interpretation of quantum mechanics, but also for improving quark model. 
}
\end{abstract}

\section{Introduction}

Without doubt, most particles can be regarded as composite particles, such
as molecules composed of atoms, atoms composed of electrons and nuclei,
nuclei composed of nucleons, so on, it is important to recognize that
physics must be invariant for the composite particles and their constituent
particles, this requirement is called particle invariance in this paper. But
difficulties arise immediately because for fermion we use Dirac equation,
for meson we use Klein-Gordon equation and for classical particle we use
Newtonian mechanics, while the connections between these equations are quite
indirect. Thus if the particle invariance is held in physics, i.e., only one
physical formalism exists for any particle, we can expect to find out the
differences between these equations by employing the particle invariance.
Using this approach is one of the goals of this paper, consequently, several
new relationships between them are found, the most important result is that
the obstacles that cluttered the path from classical mechanics to quantum
mechanics are found, it becomes possible to derive the quantum wave
equations from relativistic mechanics after the obstacles are removed.

Another goal is just to discuss interactions between particles under the
particle invariance, several new formulae of interactions are derived and
discussed. The new results provide an insight into improving quark model.

\section{Fermions and Bosons}

Fermions satisfy Fermi-Dirac statistics, Bosons satisfy Bose-Einstein
statistics, there is a connection between the spin of a particle and the
statistics. It is clear that the spin is a key concept for particle physics.
In this section we shall show that the spin of a particle is one of the
consequences of the particle invariance.

\begin{figure}[htb]
\includegraphics[bb=160 380 380 740,clip]{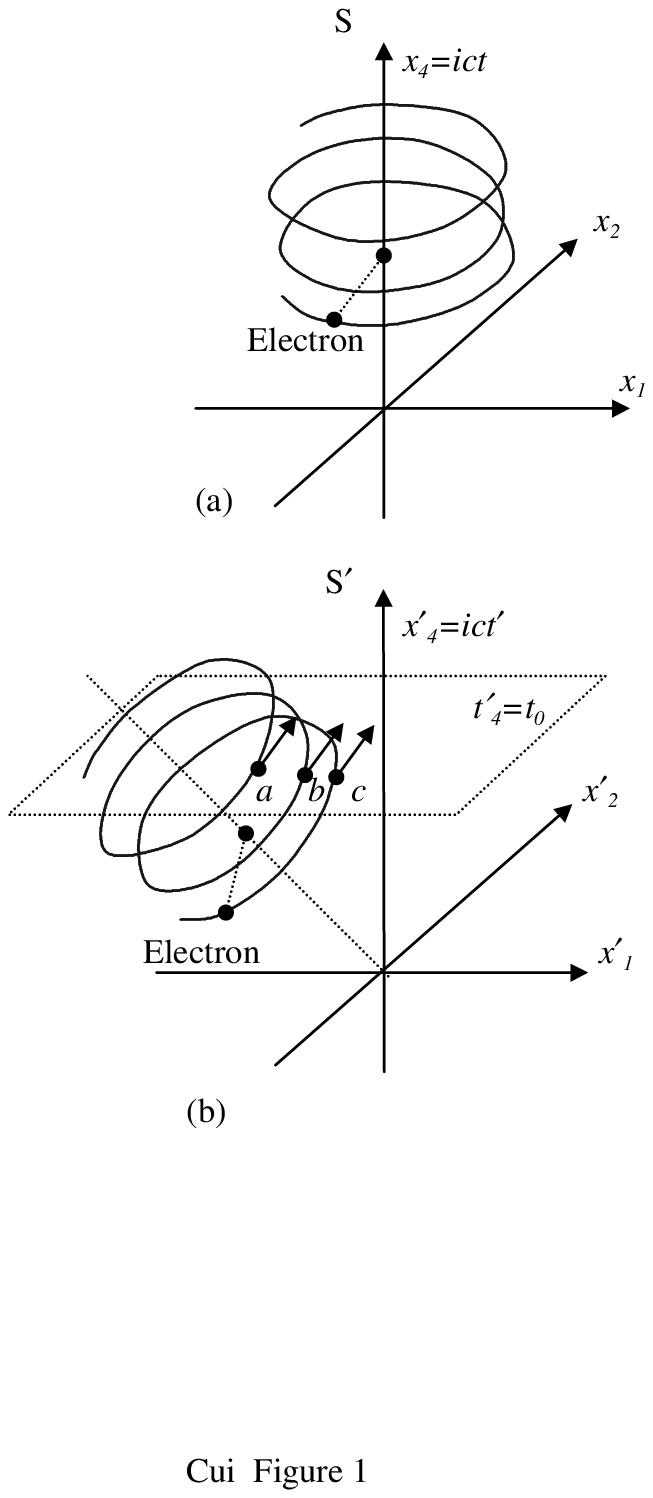}
\caption{The motion of the electron of hydrogen atom in 4-dimensional
space-time.}
\label{Dfig1}
\end{figure}

According to Newtonian mechanics, in a hydrogen atom, the single electron
revolves in an orbit about the nucleus, its motion can be described by its
position in an inertial Cartesian coordinate system $S:(x_1,x_2,x_3,x_4=ict)$%
. As the time elapses, the electron draws a spiral path (or orbit), as shown
in FIG.\ref{Dfig1}(a) in imagination.

If the reference frame $S$ ''rotates'' through an angle about the $x_2$-axis
in FIG.\ref{Dfig1}(a), becomes a new reference frame $S^{\prime }$ (there
will be a Lorentz transformation linking the frames $S$ and $S^{\prime }$),
then in the frame $S^{\prime }$, the spiral path of the electron tilts with
respect to the $x_4^{\prime }$-axis with the angle as shown in FIG.\ref
{Dfig1}(b). At one instant of time, for example, $t_4^{\prime }=t_0$
instant, the spiral path pierces many points at the plane $t_4^{\prime }=t_0$
, for example, the points labeled $a$, $b$ and $c$ in FIG.\ref{Dfig1}(b),
these points indicate that the electron can appear at many points at the
time $t_0$, in agreement with the concept of probability in quantum
mechanics. This situation gives us a hint to approach quantum wave nature
from relativistic mechanics.

Because the electron pierces the plane $t_4^{\prime }=t_0$ with 4-vector
velocity $u$, at every pierced point we can label a local 4-vector velocity
. The pierced points may be numerous if the path winds up itself into a cell
about the nucleus (due to a nonlinear effect in a sense), then the 4-vector
velocities at the pierced points form a 4-vector velocity field. It is noted
that the observation plane selected for the piercing can be taken at an
arbitrary orientation, so the 4-vector velocity field may be expressed in
general as $u(x_1,x_2,x_3,x_4)$, i.e. the velocity $u$ is a function of
4-vector position.

At every point in the reference frame $S^{\prime }$ the electron satisfies
relativistic Newton's second law of motion:

\begin{equation}
m\frac{du_\mu }{d\tau }=qF_{\mu \nu }u_\nu  \label{1}
\end{equation}
the notations consist with the convention\cite{Harris}. Since the Cartesian
coordinate system is a frame of reference whose axes are orthogonal to one
another, there is no distinction between covariant and contravariant
components, only subscripts need be used. Here and below, summation over
twice repeated indices is implied in all case, Greek indices will take on
the values 1,2,3,4, and regarding the rest mass $m$ as a constant. As
mentioned above, the 4-vector velocity $u$ can be regarded as a
multi-variable function, then

\begin{equation}
\frac{du_\mu }{d\tau }=\frac{\partial u_\mu }{\partial x_\nu }\frac{dx_\nu }{%
d\tau }=u_\nu \partial _\nu u_\mu  \label{2}
\end{equation}

\begin{equation}
qF_{\mu \nu }u_\nu =qu_\nu (\partial _\mu A_\nu -\partial _\nu A_\mu )
\label{3}
\end{equation}
Substituting them back into Eq.(\ref{1}), and re-arranging these terms, we
obtain

\begin{eqnarray}
u_\nu \partial _\nu (mu_\mu +qA_\mu ) &=&u_\nu \partial _\mu (qA_\nu ) 
\nonumber \\
&=&u_\nu \partial _\mu (mu_\nu +qA_\nu )-u_\nu \partial _\mu (mu_\nu ) 
\nonumber \\
&=&u_\nu \partial _\mu (mu_\nu +qA_\nu )-\frac 12\partial _\mu (mu_\nu u_\nu
)  \nonumber \\
&=&u_\nu \partial _\mu (mu_\nu +qA_\nu )-\frac 12\partial _\mu (-mc^2) 
\nonumber \\
&=&u_\nu \partial _\mu (mu_\nu +qA_\nu )  \label{4}
\end{eqnarray}
Using the notation

\begin{equation}
K_{\mu \nu }=\partial _\mu (mu_\nu +qA_\nu )-\partial _\nu (mu_\mu +qA_\mu )
\label{5}
\end{equation}
Eq.(\ref{4}) is given by

\begin{equation}
u_\nu K_{\mu \nu }=0  \label{6}
\end{equation}
Because $K_{\mu \nu }$ contains the variables $\partial _\mu u_\nu $, $%
\partial _\mu A_\nu $, $\partial _\nu u_\mu $ and $\partial _\nu A_\mu $
which are independent from $u_\nu $, then a main solution satisfying Eq.(\ref
{6}) is given by

\begin{eqnarray}
K_{\mu \nu } &=&0  \label{7a} \\
\partial _\mu (mu_\nu +qA_\nu ) &=&\partial _\nu (mu_\mu +qA_\mu )
\label{7b}
\end{eqnarray}
( In this paper we do not discuss the special solutions that $K_{\mu \nu
}\neq 0$, if they exist ). According to Green's formula or Stokes's theorem,
the above equation allows us to introduce a potential function $\Phi $ in
mathematics, further set $\Phi =-i\hbar \ln \psi $, we obtain a very
important equation

\begin{equation}
(mu_\mu +qA_\mu )\psi =-i\hbar \partial _\mu \psi  \label{8}
\end{equation}
where $\psi $ representing wave nature may be a complex mathematical
function, its physical meanings is determined from experiments after the
introduction of the Planck's constant $\hbar $.

The magnitude formula of 4-vector velocity of particle is given in its
square form by

\begin{equation}
-c^2=u_\mu u_\mu  \label{9}
\end{equation}
which is valid at every point in the 4-vector velocity field. Multiplying
the two sides of the above equation by $m^2\psi $ and using Eq.(\ref{8}), we
obtain

\begin{eqnarray}
-m^2c^2\psi &=&mu_\mu (-i\hbar \partial _\mu -qA_\mu )\psi  \nonumber \\
&=&(-i\hbar \partial _\mu -qA_\mu )(mu_\mu \psi )-[-i\hbar \psi \partial
_\mu (mu_\mu )]  \nonumber \\
&=&(-i\hbar \partial _\mu -qA_\mu )(-i\hbar \partial _\mu -qA_\mu )\psi 
\nonumber \\
&&-[-i\hbar \psi \partial _\mu (mu_\mu )]  \label{10}
\end{eqnarray}
According to the continuity condition for the electron motion

\begin{equation}
\partial _\mu (mu_\mu )=0  \label{11}
\end{equation}
we have

\begin{equation}
-m^2c^2\psi =(-i\hbar \partial _\mu -qA_\mu )(-i\hbar \partial _\mu -qA_\mu
)\psi  \label{12}
\end{equation}
It is known as the Klein-Gordon equation.

On the condition of non-relativity, Schrodinger equation can be derived from
the Klein-Gordon equation \cite{Schiff}(P.469).

However, we must admit that we are careless when we use the continuity
condition Eq.(\ref{11}), because, from Eq.(\ref{8}) we obtain

\begin{equation}
\partial _\mu (mu_\mu )=\partial _\mu (-i\hbar \partial _\mu \ln \psi
-qA_\mu )=-i\hbar \partial _\mu \partial _\mu \ln \psi  \label{13}
\end{equation}
where we have used Lorentz gauge condition. Thus from Eq.(\ref{9}) to Eq.(%
\ref{10}) we obtain

\begin{eqnarray}
-m^2c^2\psi &=&(-i\hbar \partial _\mu -qA_\mu )(-i\hbar \partial _\mu
-qA_\mu )\psi  \nonumber \\
&&+\hbar ^2\psi \partial _\mu \partial _\mu \ln \psi  \label{14}
\end{eqnarray}
This is of a complete wave equation for describing the motion of the
electron accurately. The Klein-Gordon equation is a linear wave equation so
that the principle of superposition is valid, however with the addition of
the last term of Eq.(\ref{14}), Eq.(\ref{14}) turns to display chaos.

In the following we shall show Dirac equation from Eq.(\ref{8}) and Eq.(\ref
{9}). From Eq.(\ref{8}), the wave function can be given in integral form by

\begin{equation}
\Phi =-i\hbar \ln \psi =\int\nolimits_{x_0}^x(mu_\mu +qA_\mu )dx_\mu +\theta
\label{15}
\end{equation}
where $\theta $ is an integral constant, $x_0$ and $x$ are the initial and
final points of the integral with an arbitrary integral path. Since
Maxwell's equations are gauge invariant, Eq.(\ref{8}) should preserve
invariant form under a gauge transformation specified by

\begin{equation}
A_\mu ^{\prime }=A_\mu +\partial _\mu \chi ,\quad \psi ^{^{\prime
}}\leftarrow \psi  \label{16}
\end{equation}
where $\chi $ is an arbitrary function. Then Eq.(\ref{15}) under the gauge
transformation is given by

\begin{eqnarray}
\psi ^{^{\prime }} &=&\exp \left\{ \frac i\hbar \int\nolimits_{x_0}^x(mu_\mu
+qA_\mu )dx_\mu +\frac i\hbar \theta \right\} \exp \left\{ \frac i\hbar
q\chi \right\}  \nonumber \\
&=&\psi \exp \left\{ \frac i\hbar q\chi \right\}  \label{17}
\end{eqnarray}
The situation in which a wave function can be changed in a certain way
without leading to any observable effects is precisely what is entailed by a
symmetry or invariant principle in quantum mechanics. Here we emphasize that
the invariance of velocity field is held for the gauge transformation.

Suppose there is a family of wave functions $\psi ^{(j)},j=1,2,3,...,N,$
which correspond to the same velocity field denoted by $P_\mu =mu_\mu $,
they are distinguishable from their different phase angles $\theta $ as in
Eq.(\ref{15}). Then Eq.(\ref{9}) can be given by

\begin{equation}
0=P_\mu P_\mu \psi ^{(j)}\psi ^{(j)}+m^2c^2\psi ^{(j)}\psi ^{(j)}  \label{18}
\end{equation}
Suppose there are four matrices $a_\mu $ which satisfy

\begin{equation}
a_{\nu lj}a_{\mu jk}+a_{\mu lj}a_{\nu jk}=2\delta _{\mu \nu }\delta _{lk}
\label{19}
\end{equation}
then Eq.(\ref{18}) can be rewritten as

\begin{eqnarray}
0 &=&a_{\mu kj}a_{\mu jk}P_\mu \psi ^{(k)}P_\mu \psi ^{(k)}  \nonumber \\
&&+(a_{\nu lj}a_{\mu jk}+a_{\mu lj}a_{\nu jk})P_\nu \psi ^{(l)}P_\mu \psi
^{(k)}|_{\nu \geq \mu ,when\nu =\mu ,l\neq k}  \nonumber \\
&&+mc\psi ^{(j)}mc\psi ^{(j)}  \nonumber \\
&=&[a_{\nu lj}P_\nu \psi ^{(l)}+i\delta _{lj}mc\psi ^{(l)}][a_{\mu jk}P_\mu
\psi ^{(k)}-i\delta _{jk}mc\psi ^{(k)}]  \nonumber \\
&&  \label{20}
\end{eqnarray}
Where $\delta _{jk}$ is Kronecker delta function, $j,k,l=1,2,...,N$. For the
above equation there is a special solution given by

\begin{equation}
\lbrack a_{\mu jk}P_\mu -i\delta _{jk}mc]\psi ^{(k)}=0  \label{21}
\end{equation}

There are many solutions for $a_\mu $ which satisfy Eq.(\ref{19}), we select
a set of $a_\mu $ as

\begin{eqnarray}
N &=&4,\quad a_\mu =\gamma _\mu \quad (\mu =1,2,3,4)  \label{22a} \\
\gamma _n &=&-i\beta \alpha _n\quad (n=1,2,3),\quad \gamma _4=\beta
\label{22b}
\end{eqnarray}
where $\gamma _\mu ,\alpha $ and $\beta $ are the matrices defined in Dirac
algebra\cite{Harris}(P.557). Substituting them into Eq.(\ref{21}), we obtain

\begin{equation}
\lbrack ic(-i\hbar \partial _4-qA_4)+c\alpha _n(-i\hbar \partial
_n-qA_n)+\beta mc^2]\psi =0  \label{23}
\end{equation}
where $\psi $ is an one-column matrix about $\psi ^{(k)}$. Then Eq.(\ref{23}%
) is just the Dirac equation.

The Dirac equation is a linear wave equation, the principle of superposition
is valid for it. Let index $s$ denote velocity field, then $\psi _s(x)$,
whose four component functions correspond to the same velocity field $s$,
may be regarded as the eigenfunction of the velocity field $s$ ( it may be
different from the eigenfunction of energy ). Because the velocity field is
an observable in a physical system, in quantum mechanics we know, $\psi
_s(x) $ constitute a complete basis in which arbitrary function $\phi (x)$
can be expanded in terms of them

\begin{equation}
\phi (x)=\int C(s)\psi _s(x)ds  \label{24}
\end{equation}
Obviously, $\phi (x)$ satisfies Eq.(\ref{23}). Then Eq.(\ref{23}) is just
the Dirac equation suitable for composite wave function.

Alternatively, another method to show the Dirac equation is more
traditional: At first, we show the Dirac equation of free particle by
employing plane waves, we easily obtain Eq.(\ref{23}) on the condition of $%
A_\mu =0$; Next, adding electromagnetic field, the plane waves are still
valid in any finite small volume with the momentum of Eq.(\ref{8}) when we
regard the field to be uniform in the volume, so the Dirac equation Eq.(\ref
{23}) is valid in the volume even if $A_\mu \neq 0$, the plane waves
constitute a complete basis in the volume; Third, the finite small volume
can be chosen to locate at anywhere, then anywhere have the same complete
basis, therefore the Dirac equation Eq.(\ref{23}) is valid at anywhere.

Of course, on the condition of non-relativity, Schrodinger equation can be
derived from the Dirac equation \cite{Schiff}(P.479).

By further calculation, the Dirac equation can arrive at Klein-Gordon
equation with an additional term which represents the effect of spin, this
term is just the last term of Eq.(\ref{14}) approximately.

But, do not forget that the Dirac equation is a special solution of Eq.(\ref
{20}), therefore we believe there are some quantum effects beyond the Dirac
equation. With this consequence, it is easy to understand why some problems
of quantum electrodynamics can not been completely explained by the Dirac
equation.

Eq.(\ref{20}) originates from the magnitude formula of 4-vector velocity of
particle, the formula is suitable for any particle, so it satisfies the
particle invariance. The Dirac equation is regarded as an approximation to
Eq.(\ref{20}), the approximation brings out many troubles with the spin
concept. From the Dirac equation we can predict that a composite particle
and an its constituent both have their own spins, but this prediction is not
true for mesons because Pion has zero spin while its constituent quark has
1/2 spin, in other words, due to the approximation the Dirac equation does
not involve some states such as zero spin state. That is why we want to
classify particles into fermions and mesons by spin and use different
equations. If we can find a precise solution of Eq.(\ref{20}) instead of the
Dirac equation, then the classification is not necessary. It is noted that
Eq.(\ref{20}) is nonlinear while the Dirac equation is linear, this
reminders us that we can never find any precise solutions in a linear
equation which satisfy Eq.(\ref{20}). Therefore, for this problem, a good
solution depends on how much precision we can reach for our requirement.

In one hand, it is rather remarkable that Klein-Gordon equation and Dirac
equation can be derived from relativistic Newton's second law of motion
approximately, in another hand, all particles, such as fermions, bosons and
classical particles, satisfy the relativistic Newton's second law (it will
be further clear later), thus it is a natural choice that only the
relativistic Newton's second law is independent and necessary. Only one
formalism is necessary for any particle, this is just the particle
invariance, we arrive at the aim.

As mentioned above, the spin is one feature hidden in the relativistic
Newton's second law, but more features will turn out from the relativistic
Newton's second law in the following sections.

\section{Determining the Planck's constant}

In this section we discuss how to determine the Planck's constant that
emerges in the preceding section.

In 1900, M. Planck assumed that the energy of a harmornic oscilator can take
on only discrete values which are integral multiples of $h\nu $, where $\nu $
is the vibration frequency and $h$ is a fundamental constant, now either $h$
or $\hbar =h/2\pi $ is called as Planck's constant. The Planck's constant
next made its appearance in 1905, when Einstein used it to explain the
photoelectric effect, he assumed that the energy in an electromagnetic wave
of frequency $\omega $ is in the form of discrete quanta (photons) each of
which has an energy $\hbar \omega $ in accordance with Planck's assumption.
From then, it has been recognized that the Planck's constant plays a key
role in quantum mechanics.

According to the previous section, no mater how to move or when to move in
Minkowski's space, the motion of a particle is governed by a potential
function $\Phi $\ as

\begin{equation}
mu_\mu +qA_\mu =\partial _\mu \Phi  \label{p1}
\end{equation}
For applying Eq.(\ref{p1}) to specific applications,without loss of
generality, we set $\Phi =-i\kappa \psi $, then Eq.(\ref{p1}) is rewritten as

\begin{equation}
(mu_\mu +qA_\mu )\psi =-i\kappa \partial _\mu \psi  \label{p2}
\end{equation}
the coefficient $\kappa $\ is subject to the interpretation of $\psi $.

There are three mathematical properties of $\psi $ worth recording here.
First, if there is a path $l_i$ joining initial point $x_0$ to final point $%
x $, then

\begin{equation}
\psi _i=e^{\frac i\kappa \int\nolimits_{x_0(l_i)}^x(mu_\mu +qA_\mu )dx_\mu }
\label{p3}
\end{equation}
Second, the integral of Eq.(\ref{p3}) is independent from the choice of
path. Third, the superposition principle is valid for $\psi _i$, i.e., if
there are $N$ paths from $x_0$ to $x$, then

\begin{equation}
\psi =\sum\limits_i^N\psi _i  \label{p4}
\end{equation}

\begin{equation}
m\overline{u_\mu }=\sum\limits_i^Nmu_\mu \psi _i/\sum\limits_i^N\psi _i\quad
\label{p5}
\end{equation}

\begin{equation}
(m\overline{u_\mu }+qA_\mu )\psi =-i\kappa \partial _\mu \psi  \label{p6}
\end{equation}
where $m\overline{u_\mu }$ is the average momentum.

To gain further insight into physical meanings of this equations, we shall
discuss two applications.

\subsection{Two slit experiment}

\begin{figure}[ht]
\includegraphics[bb=110 540 310 720,clip]{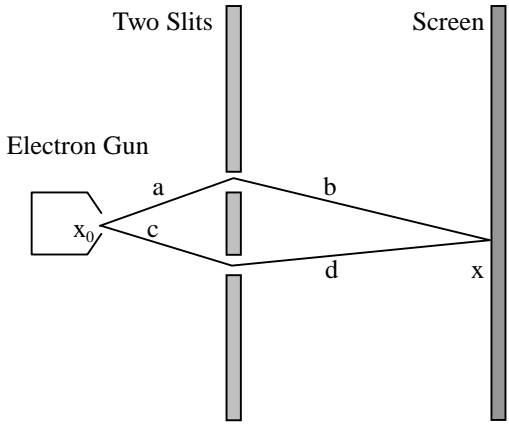}
\caption{A diffraction experiment in which electron beam from the gun
through the two slits to form a diffractin pattern at the screen.}
\label{Pfig1}
\end{figure}

As shown in FIG.\ref{Pfig1}, suppose that the electron gun emits a burst of
electrons at $x_0$ at time $t=0$, the electrons arrive at the point $x$ on
the screen at time $t$. There are two paths for the electron to go to the
destination, according to our above statement, $\psi $ is given by

\begin{equation}
\psi =e^{\frac i\kappa \int\nolimits_{x_0(l_1)}^x(mu_\mu )dx_\mu }+e^{\frac
i\kappa \int\nolimits_{x_0(l_2)}^x(mu_\mu )dx_\mu }  \label{p7}
\end{equation}
where we use $l_1$ and $l_2$ to denote the paths $a+b$ and $c+d$
respectively. Multiplying Eq.(\ref{p7}) by its complex conjugate gives

\begin{eqnarray}
W &=&\psi (x)\psi ^{*}(x)  \nonumber \\
&=&2+e^{\frac i\kappa \int\nolimits_{x_0(l_1)}^x(mu_\mu )dx_\mu -\frac
i\kappa \int\nolimits_{x_0(l_2)}^x(mu_\mu )dx_\mu }  \nonumber \\
&&+e^{\frac i\kappa \int\nolimits_{x_0(l_2)}^x(mu_\mu )dx_\mu -\frac i\kappa
\int\nolimits_{x_0(l_1)}^x(mu_\mu )dx_\mu }  \nonumber \\
&=&2+2\cos [\frac 1\kappa \int\nolimits_{x_0(l_1)}^x(mu_\mu )dx_\mu 
\nonumber \\
&&-\frac 1\kappa \int\nolimits_{x_0(l_2)}^x(mu_\mu )d_\mu ]  \nonumber \\
&=&2+2\cos [\frac p\kappa (l_1-l_2)]  \label{p8}
\end{eqnarray}
where $p$ is the momentum of the electron. We find a typical interference
pattern with constructive interference when $l_1-l_2$ is an integral
multiple of $\kappa /p$, and destructive interference when it is a half
integral multiple. This kind of experiments has been done since a long time
age, no mater what kind of particle, the comparison of the experiments to
Eq.(\ref{p8}) leads to two consequences: (1) the complex function $\psi $ is
found to be probability amplitude, i.e., $\psi (x)\psi ^{*}(x)$ expresses
the probability of finding a particle at location $x$ in the Minkowski's
space. (2) $\kappa $ is the Planck's constant.

The integral of time component in the above calculation has been
autimatically canceled because the experimental pattern is stable.

\subsection{The Aharonov-Bohm effect}

\begin{figure}[ht]
\includegraphics[bb=110 540 310 720,clip]{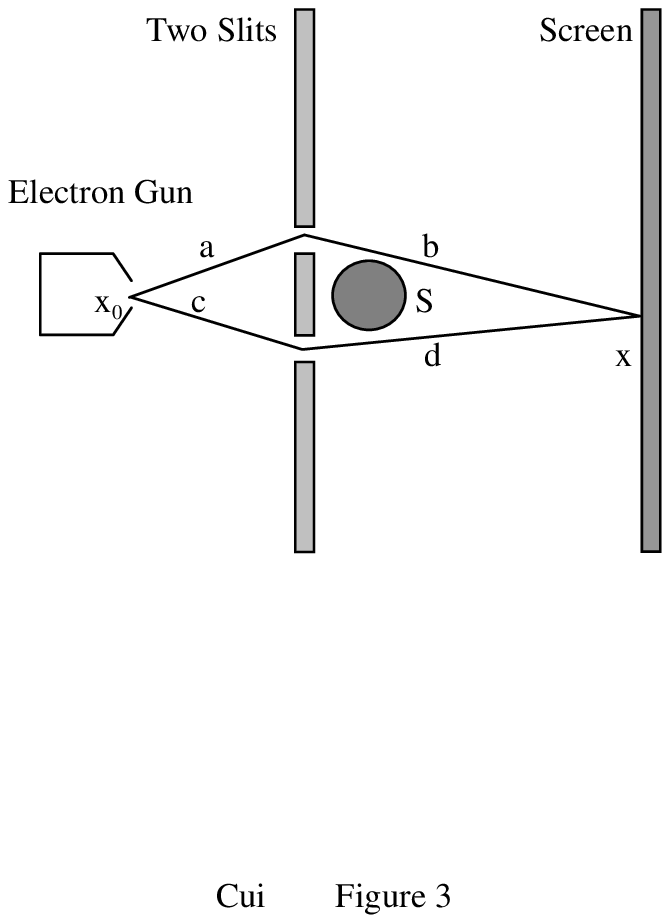}
\caption{A diffraction experiment with adding a solenoid.}
\label{Pfig2}
\end{figure}

Let us consider the modification of the two slit experiment, as shown in FIG.%
\ref{Pfig2}. Between the two slits there is located a tiny solenoid S,
designed so that a magnetic field perpendicular to the plane of the figure
can be produced in its interior. No magnetic field is allowed outside the
solenoid, and the walls of the solenoid are such that no electron can
penetrate to the interior. Like Eq.(\ref{p7}), the amplitude $\psi $ is
given by

\begin{equation}
\psi =e^{\frac i\kappa \int\nolimits_{x_0(l_1)}^x(mu_\mu +qA_\mu )dx_\mu
}+e^{\frac i\kappa \int\nolimits_{x_0(l_2)}^x(mu_\mu +qA_\mu )dx_\mu }
\label{p9}
\end{equation}
and the probability is given by

\begin{eqnarray}
W&=&\psi (x)\psi ^{*}(x)  \nonumber \\
&=&2+e^{\frac i\kappa \int\nolimits_{x_0(l_1)}^x(mu_\mu +qA_\mu )dx_\mu
-\frac i\kappa \int\nolimits_{x_0(l_2)}^x(mu_\mu +qA_\mu )dx_\mu }  \nonumber
\\
&&+e^{\frac i\kappa \int\nolimits_{x_0(l_2)}^x(mu_\mu +qA_\mu )dx_\mu -\frac
i\kappa \int\nolimits_{x_0(l_1)}^x(mu_\mu +qA_\mu )dx_\mu }  \nonumber \\
&=&2+2\cos [\frac p\kappa (l_1-l_2)+\frac 1\kappa
\int\nolimits_{x_0(l_1)}^xqA_\mu dx_\mu  \nonumber \\
&&-\frac 1\kappa \int\nolimits_{x_0(l_2)}^xqA_\mu dx_\mu ]  \nonumber \\
&=&2+2\cos [\frac p\kappa (l_1-l_2)+\frac 1\kappa \oint_{(l_1+\overline{l_2}%
)}qA_\mu dx_\mu ]  \nonumber \\
&=&2+2\cos [\frac p\kappa (l_1-l_2)+\frac{q\phi }\kappa ]  \label{p10}
\end{eqnarray}
where $\overline{l_2}$ denotes the inverse path to the path $l_2$, $\phi $
is the magnetic flux that passes through the surface between the paths $l_1$
and $\overline{l_2}$, and it is just the flux inside the solenoid.

Now, constructive (or destructive) interference occurs when

\begin{equation}
\frac p\kappa (l_1-l_2)+\frac{q\phi }\kappa =2\pi n\quad (or\quad n+\frac 12)
\label{p11}
\end{equation}
where $n$ is an integer. When $\kappa $ takes the value of the Planck's
constant, we know that this effect is just the Aharonov-Bohm effect which
was shown experimentally in 1960.

\section{The directions of forces}

\label{direction}

In this section we shall correct a mistake about Coulomb's force and
gravitational force in physical education, which cluttered the path from
classical mechanics to quantum mechanics. We also shall discuss Maxwell's
equations in detail .

In the world, almost every young person was educated to know that the
Coulomb's force and gravitational force act along the line joining a couple
of particles, but this knowledge is incorrect in the theory of relativity .

In relativity theory, the 4-vector velocity $u$ of a particle has components 
$u_\mu $, the magnitude of the 4-vector velocity $u$ is given by

\begin{equation}
|u|=\sqrt{u_\mu u_\mu }=\sqrt{-c^2}=ic  \label{e01}
\end{equation}
The above equation is valid so that any force can never change $u$ in the
magnitude but can change $u$ in the direction. We therefore conclude that
the Coulomb's force and gravitational force on a particle always act in the
direction orthogonal to the 4-vector velocity of the particle in the
4-dimensional space-time, rather than along the line joining a couple of
particles. Alternatively, any 4-vector force $f$ satisfy the following
perpendicular or orthogonal relation

\begin{equation}
u_\mu f_\mu =u_\mu m\frac{du_\mu }{d\tau }=\frac m2\frac{d(u_\mu u_\mu )}{%
d\tau }=0  \label{e02}
\end{equation}

This simple inference clearly tells us that the forces are not centripetal
or centrifugal forces about their sources, even if in 3-dimensional space
[see Eq.(\ref{e7}) ], this character provides a internal reason for
accounting for the quantum behavior of particle or chaos. Thus the
derivations in terms of 4-vector velocity field in the preceding section
become reasonable.

In the present paper, Eq.(\ref{e02}) has been elevated to an essential
requirement for definition of force, which brings out many new aspects for
Coulomb's force and gravitational force.

\subsection{Coulomb's force and Lorentz force}

\begin{figure}[htb]
\includegraphics[bb=175 580 355 740,clip]{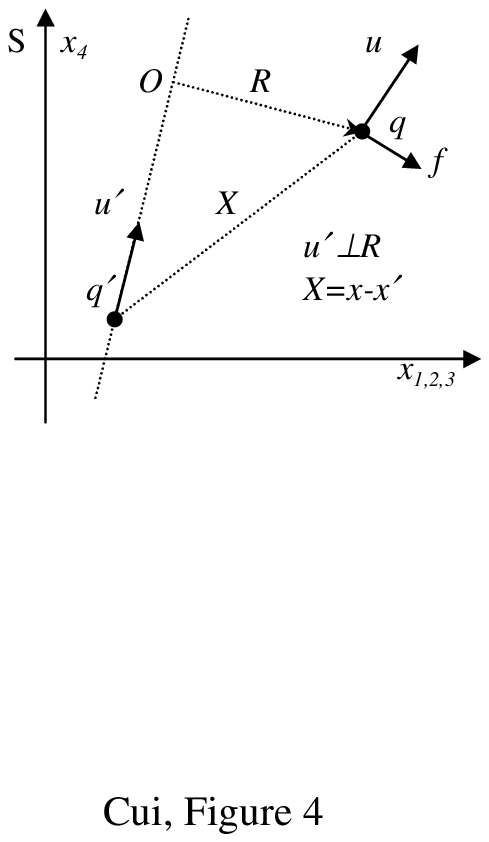}
\caption{The Coulomb's force acting on $q$ is perpendicular to the 4-vector
velocity $u$ of $q$, and lies in the plane of $u^{\prime }$ and $X$ with the
retardation with respect to $q^{\prime }$.}
\label{Afig1}
\end{figure}

We assume that Coulomb's law remains valid only for two particles both at
rest in usual 3-dimensional space. Suppose there are two charged particle $q$
and $q^{\prime }$ locating at positions $x$ and $x^{\prime }$ in a Cartesian
coordinate system $S$ and moving at 4-vector velocities $u$ and $u^{\prime }$
respectively, as shown in FIG.\ref{Afig1}, where we use $X$ to denote $%
x-x^{\prime }$. The Coulomb's force $f$ acting on particle $q$ is
perpendicular (orthogonal) to the velocity direction of $q$, as illustrated
in FIG.\ref{Afig1}, like a centripetal force, the force $f$ should make an
attempt to rotate itself about its path center, the center may locate at the
front or back of the particle $q^{\prime }$, so the force $f$ should lie in
the plane of $u^{\prime }$ and $X$, then

\begin{equation}
f=Au^{\prime }+BX  \label{e1}
\end{equation}
where $A$ and $B$ are unknown coefficients, the possibility of this
expansion will be further clear in the next subsection in where the
expansion is not an assumption [see Eq.(\ref{g3})]. Using the relation $%
f\perp u$, we get

\begin{equation}
u\cdot f=A(u\cdot u^{\prime })+B(u\cdot X)=0  \label{e2}
\end{equation}
we rewrite Eq.(\ref{e1}) as

\begin{equation}
f=\frac A{u\cdot X}[(u\cdot X)u^{\prime }-(u\cdot u^{\prime })X]  \label{e3}
\end{equation}
It follows from the direction of Eq.(\ref{e3}) that the unit vector of the
Coulomb's force direction is given by

\begin{equation}
\widehat{f}=\frac 1{c^2r}[(u\cdot X)u^{\prime }-(u\cdot u^{\prime })X]
\label{e4}
\end{equation}
because

\begin{eqnarray}
\widehat{f} &=&\frac 1{c^2r}[(u\cdot X)u^{\prime }-(u\cdot u^{\prime })X] 
\nonumber \\
&=&\frac 1{c^2r}[(u\cdot R)u^{\prime }-(u\cdot u^{\prime })R]  \nonumber \\
&=&-[(\widehat{u}\cdot \widehat{R})\widehat{u}^{\prime }-(\widehat{u}\cdot 
\widehat{u}^{\prime })\widehat{R}]  \nonumber \\
&=&-\widehat{u}^{\prime }\cosh \alpha +\widehat{R}\sinh \alpha  \label{e5a}
\end{eqnarray}

\begin{equation}
|\widehat{f}|=1  \label{e5b}
\end{equation}
Where $\alpha $ refers to the angle between $u$ and $R$, $R\perp u^{\prime }$%
, $r=|R|$, $\widehat{u}=u/ic$, $\widehat{u}^{\prime }=u^{\prime }/ic$, $%
\widehat{R}=R/r$. Suppose that the magnitude of the force $f$ has classical
form

\begin{equation}
|f|=k\frac{qq^{\prime }}{r^2}  \label{e6}
\end{equation}
Combination of Eq.(\ref{e6}) with (\ref{e4}), we obtain a modified Coulomb's
force

\begin{eqnarray}
f &=&\frac{kqq^{\prime }}{c^2r^3}[(u\cdot X)u^{\prime }-(u\cdot u^{\prime })X%
]  \nonumber \\
&=&\frac{kqq^{\prime }}{c^2r^3}[(u\cdot R)u^{\prime }-(u\cdot u^{\prime })R]
\label{e7}
\end{eqnarray}
This force is in the form of Lorentz force for the two particles, relating
with the Ampere's law and Biot-Savart-Laplace law.

It follows from Eq.(\ref{e7}) that the force can be rewritten in terms of
4-vector components as

\begin{eqnarray}
f_\mu &=&qF_{\mu \nu }u_\nu  \label{e8a} \\
F_{\mu \nu } &=&\partial _\mu A_\nu -\partial _\nu A_\mu  \label{e8b} \\
A_\mu &=&\frac{kq^{\prime }}{c^2}\frac{u_\mu ^{\prime }}r  \label{e8c}
\end{eqnarray}
Where we have used the relations

\begin{equation}
\partial _\mu \left( \frac 1r\right) =-\frac {R_\mu }{r^3}  \label{e9}
\end{equation}

\subsection{Lorentz gauge condition}

From Eq.(\ref{e8c}), because of $u^{\prime }\perp R$ , we have

\begin{equation}
\partial _\mu A_\mu =\frac{kq^{\prime }u_\mu ^{\prime }}{c^2}\partial _\mu
\left( \frac 1r\right) =-\frac{kq^{\prime }u_\mu ^{\prime }}{c^2}\left( 
\frac{R_\mu }{r^3}\right) =0  \label{e11}
\end{equation}
It is known as the Lorentz gauge condition.

\subsection{Maxwell's equations}

To note that $R$ has three degrees of freedom on the condition $R\perp
u^{\prime }$, so we have

\begin{equation}
\partial _\mu R_\mu =3  \label{e12}
\end{equation}

\begin{equation}
\partial _\mu \partial _\mu \left( \frac 1r\right) =-4\pi \delta (R)
\label{e13}
\end{equation}
From Eq.(\ref{e8b}), we have

\begin{eqnarray}
\partial _\nu F_{\mu \nu } &=&\partial _\nu \partial _\mu A_\nu -\partial
_\nu \partial _\nu A_\mu =-\partial _\nu \partial _\nu A_\mu  \nonumber \\
&=&-\frac{kq^{\prime }u_\mu ^{\prime }}{c^2}\partial _\nu \partial _\nu
\left( \frac 1r\right) =\frac{kq^{\prime }u_\mu ^{\prime }}{c^2}4\pi \delta
(R)  \nonumber \\
&=&\mu _0J_\nu ^{\prime }  \label{e14}
\end{eqnarray}
where we define $J_\nu ^{\prime }=q^{\prime }u_\nu ^{\prime }\delta (R)$.
From Eq.(\ref{e8b}), by exchanging the indices and taking the summation of
them, we have

\begin{equation}
\partial _\lambda F_{\mu \nu }+\partial _\mu F_{\nu \lambda }+\partial _\nu
F_{\lambda \mu }=0  \label{e15}
\end{equation}
The Eq.(\ref{e14}) and (\ref{e15}) are known as the Maxwell's equations. For
continuous media, they are valid as well.

\subsection{Lienard-Wiechert potential}

From the Maxwell's equations, we know there is a retardation time for action
to propagate between the two particles, the retardation effect is measured by

\begin{equation}
r=c\Delta t=c\frac{\overline{q^{\prime }O}}{ic}=c\frac{\widehat{u}^{\prime
}\cdot X}{ic}=\frac{u_\nu ^{\prime }(x_\nu ^{\prime }-x_\nu )}c  \label{e17}
\end{equation}
as illustrated in FIG.\ref{Afig1}. Then

\begin{equation}
A_\mu =\frac{kq^{\prime }}{c^2}\frac{u_\mu ^{\prime }}r=\frac{kq^{\prime }}c%
\frac{u_\mu ^{\prime }}{u_\nu ^{\prime }(x_\nu ^{\prime }-x_\nu )}
\label{e18}
\end{equation}

Obviously, Eq.(\ref{e18}) is known as the Lienard-Wiechert potential for a
moving particle.

The above formalism clearly shows that Maxwell's equations can be derived
from the classical Coulomb's force and the perpendicular (orthogonal)
relation of force and velocity. In other words, the perpendicular relation
is hidden in Maxwell's equation. Specially, Eq.(\ref{e3}) directly accounts
for the geometrical meanings of curl of vector potential, the curl contains
the perpendicular relation. Since the perpendicular relation of force and
velocity is one of the consequences from relativistic Newton's second law,
it is also one of the features from the particle invariance.

\subsection{Gravitational force}

The above formalism has a significance on guiding how to develope the theory
of gravity. In analogy with the modified Coulomb's force of Eq.(\ref{e7}),
we directly suggest a modified universal gravitational force as

\begin{eqnarray}
f &=&-\frac{Gmm^{\prime }}{c^2r^3}[(u\cdot X)u^{\prime }-(u\cdot u^{\prime
})X]  \nonumber \\
&=&-\frac{Gmm^{\prime }}{c^2r^3}[(u\cdot R)u^{\prime }-(u\cdot u^{\prime })R]
\label{g1}
\end{eqnarray}
for a couple of particles with masses $m$ and $m^{\prime }$ respectively.

Comparing with some incorrect statements about Coulomb's force and
gravitational force in most textbooks, and for emphasizing its feature, the
perpendicular (orthogonal) relation of force and velocity was called the
direction adaptation nature of force in the author's previous paper\cite
{Cui0173}.

\subsection{The Magnet-like components of the gravitational force}

We emphasize that the perpendicular relation of force and velocity must be
valid if gravitational force can be defined as a force. It follows from Eq.(%
\ref{g1}) that we can predict that there are gravitational radiation and
magnet-like components for the gravitational force. Particularly, the
magnet-like components will act as a key role in the geophysics and
atmosphere physics.

If we have not any knowledge but know there exists the classical universal
gravitation $\mathbf{f}$ between two particles $m$ and $m^{\prime }$, what
form will take the 4-vector gravitational force $f$ ? Suppose that $%
m^{\prime }$ is at rest at the origin, using $u=(\mathbf{u},u_4)$, $%
u^{\prime }=(0,0,0,ic)$ and $u\cdot f=0$, we have

\begin{equation}
f_4=\frac{u_4f_4}{u_4}=\frac{u\cdot f-\mathbf{u}\cdot \mathbf{f}}{u_4}=-%
\frac{\mathbf{u}\cdot \mathbf{f}}{u_4}  \label{g2}
\end{equation}

\begin{eqnarray}
f &=&(\mathbf{f},f_4)=\mathbf{f}+f_4\frac{u^{\prime }}{ic}=\mathbf{f}-(\frac{%
\mathbf{u}\cdot \mathbf{f}}{u_4})\frac{u^{\prime }}{ic}  \nonumber \\
&=&\frac 1{icu_4}[icu_4\mathbf{f}-(\mathbf{u}\cdot \mathbf{f})u^{\prime }] 
\nonumber \\
&=&\frac 1{icu_4}[(u^{\prime }\cdot u)\mathbf{f}-(\mathbf{u}\cdot \mathbf{f}%
)u^{\prime }]  \nonumber \\
&=&\frac{|\mathbf{f}|}{icu_4|\mathbf{R}|}[(u^{\prime }\cdot u)\mathbf{R}-(%
\mathbf{u}\cdot \mathbf{R})u^{\prime }]  \nonumber \\
&=&\frac{|\mathbf{f}|}{icu_4|\mathbf{R}|}[(u^{\prime }\cdot u)R-(u\cdot
R)u^{\prime }]  \label{g3}
\end{eqnarray}
where $R\perp u^{\prime }$, $R=(\mathbf{R},0)$. If we ''rotate'' our frame
of reference to make $m^{\prime }$ not to be at rest, Eq.(\ref{g3}) will
still be valid because of covariance. Then we find the 4-vector
gravitational force goes back to the form of Eq.(\ref{g1}), like Lorentz
force, having the magnet-like components.

It is noted that the perpendicular relation of force and velocity is valid
for any force: strong, electromagnetic, weak and gravitational interactions,
therefore there are many new aspects remaining for physics to explore.

\section{Interactions between particles}

Under the invariance of particle, the most simple model of particle is that
all particles are composed of identical constituents, the constituent is
regarded as ''the most elementary and most small particle'' in the world.
Since quarks have never been observed, our speculation leads us to propose a
better model to organize known data. For this challenging purpose, in the
present paper, we introduce a fictitious elementary particle, given a name
''Dollon'' for our convenience, to assemble other particles such as
fermions, mesons or classical particles, the Dollon is regarded as ''the
most elementary and most small particle'' in the world. Our work focuses on
conceptual development.

\subsection{Basic force}

\label{basicf}Consider a Dollon moving in Minkowski's space $%
(x_1,x_2,x_3,x_4=ict)$ with 4-vector velocity $u=(\mathbf{u},u_4)$, the
motion of the Dollon satisfies the magnitude formula of 4-vector velocity of
particle

\begin{equation}
u_\mu u_\mu =-c^2  \label{r1}
\end{equation}
Differentiating the above equation with respect to the proper time interval $%
d\tau $ of the Dollon gives

\begin{equation}
\frac{d\mathbf{u}}{d\tau }=\mathbf{f}\qquad \qquad \frac{du_4}{d\tau }=-%
\frac{\mathbf{u}\cdot \mathbf{f}}{u_4}  \label{r2}
\end{equation}
where the result has been written in the two parts by defining a
3-dimensional vector $\mathbf{f}$. Defining a 4-vector

\begin{equation}
f=(\mathbf{f},-\frac{\mathbf{u}\cdot \mathbf{f}}{u_4})  \label{r3}
\end{equation}
then from Eq.(\ref{r1}) we have readily

\begin{equation}
\frac{du}{d\tau }=f\qquad \qquad u\cdot f=u_\mu f_\mu =0  \label{r4}
\end{equation}
It means that $u$ and $f$ are orthogonal with each other.

Consider two particles ''Bob'' and ''Alice'' located at $x$ and $x^{\prime }$
in the 4-dimensional space respectively, they are composed of many Dollons,
the number of Dollons in Alics is $M$, and in Bob is $m$, when Bob and Alice
move with 4-vector velocities $u$ and $u^{\prime }$ respectively, following
Eq.(\ref{r2}), they can be assigned two sets of motion equations as

\begin{equation}
Bob:m\frac{d\mathbf{u}}{d\tau }=m\mathbf{f}\qquad \quad m\frac{du_4}{d\tau }%
=-m\frac{\mathbf{u}\cdot \mathbf{f}}{u_4}  \label{r5}
\end{equation}

\begin{equation}
Alice:M\frac{d\mathbf{u}^{\prime }}{d\tau ^{\prime }}=M\mathbf{f}^{\prime
}\qquad \quad M\frac{du_4^{\prime }}{d\tau ^{\prime }}=-M\frac{\mathbf{u}%
^{\prime }\cdot \mathbf{f}^{\prime }}{u_4^{\prime }}  \label{r6}
\end{equation}

Now we have a question: what is the interaction between Bob and Alice?
Obviously, the form of Eq.(\ref{r5}) seems to be relativistic Newton second
law for Bob, $\mathbf{f}$ seems to be a 3-vector force, $\mathbf{u}\cdot 
\mathbf{f}$ seems to be the rate at which the force does work on Bob. For
seeking for further answers, we need to recall the Newton's first law of
motion, the law is valid in theory of relativity and reads

\textit{First Law: An object at rest will remain at rest and an object in
motion will continue to move in a straight line at constant speed forever
unless some net external force acts to change this motion.}

If the object is a composite system composed of many Dollons, then we can
understand the First Law with three consequences.

Consequence 1: Let $S$ denote the number of Dollons in a composite system,
the average velocity of the system is defined as

\begin{equation}
u_c=\frac 1S\sum\limits_i^Su^{(i)}  \label{r7}
\end{equation}
where $u^{(i)}$ is the 4-vector velocity of the ith Dollon. The average
velocity represents the motion of the center of the system. The Fist Law
only means that the center of the system remains at rest or in motion, i.e.,
rotation about its center is permited.

Consequence 2: The total number of Dollons in the system must be unchanged,
i.e., the conservation of Dollon number must be held, otherwise any creation
or annihilation of Dollon will lead to a sudden shift of the center of the
system.

Consequence 3: When two bodies are seperated from an infinite distance, the
interaction between them must vanish. Otherwise, no body can be at rest,
bacause a rest body will always be affected by the motion of a far distance
body, whereas the far distance bodies are innumerable as a background.

Now we go back to consider the whole system composed of Bob and Alics,
without loss of generality, suppose that the center is at rest at the origin
of the frame of reference, then the center has a 4-vector velocity $%
u_c=(0,0,0,u_{4c})$, the ''at rest'' refers to being at rest in usual
3-dimensional space. From Eq.(\ref{r4}), the quantity $f$ must be orthogonal
with the 4-vector velocity $u$ of Bob, likewise for Alice, we have

\begin{eqnarray}
Bob &:&\qquad u\cdot f=u_\mu f_\mu =0  \label{r8} \\
Alice &:&\qquad u^{\prime }\cdot f^{\prime }=u_\mu ^{\prime }f_\mu ^{\prime
}=0  \label{r9}
\end{eqnarray}
They set up a rule for the interaction between Bob and Alice in the
composite system. We specially choose to study the interaction which happens
at such instant that the position vector $X$ of Bob with respect with Alice
(i.e., $X=x-x^{\prime }$) is orthogonal to $u$ and $u^{\prime }$
simultaneously.

\begin{eqnarray}
Bob &:&\qquad u\cdot X=0  \label{r91} \\
Alice &:&\qquad u^{\prime }\cdot X=0  \label{r92}
\end{eqnarray}
The existence of such instant of time will become clear in the subsection 
\ref{X}. From Eq.(\ref{r8}) and Eq.(\ref{r9}), we get parallel relations

\begin{eqnarray}
Bob &:&\qquad f\propto X  \label{r10} \\
Alice &:&\qquad f^{\prime }\propto X  \label{r11}
\end{eqnarray}
For Bob, using notation $X=(\mathbf{r},X_4)$, $r=|\mathbf{r}|$,
vector-multiplying Eq.(\ref{r5}) by $\mathbf{r}$, because $\mathbf{f}$
parallels $\mathbf{r}$, we have

\begin{equation}
\mathbf{r\times (}m\frac{d\mathbf{u}}{d\tau })=m\frac{d(\mathbf{r}\times 
\mathbf{u})}{d\tau }=m\mathbf{r}\times \mathbf{f=0}  \label{r12}
\end{equation}
It means

\begin{equation}
\mathbf{r}\times \mathbf{u}=\mathbf{h}=const.  \label{r13}
\end{equation}
where $\mathbf{h}$ is an integral constant. Likewise for Alice. From Eq.(\ref
{r10}) we can expand $\mathbf{f}/u_4$ in a Taylor series in $1/r$, this gives

\begin{equation}
\frac{\mathbf{f}}{u_4}=\frac{\mathbf{r}}r(b_0+b_1\frac 1r+b_2\frac
1{r^2}+b_3\frac 1{r^3}+...)  \label{r14}
\end{equation}

From Eq.(\ref{r5}) we obtain

\begin{eqnarray}
u_4 &=&\int (-\frac{\mathbf{u}\cdot \mathbf{f}}{u_4})d\tau =-\int \frac{|%
\mathbf{f|}}{u_4}dr  \nonumber \\
&=&\varepsilon -b_0r-b_1\ln r+b_2\frac 1r+b_3\frac 1{2r^2}+...  \label{r15}
\end{eqnarray}
where $\varepsilon $ is an integral constant. Now consider Eq.(\ref{r13}),
it means that Bob moves around Alice (no matter by attractive or repulsive
interaction), when $h\rightarrow 0$, Bob may access Alice as close as
possible at perihelion point, at the perihelion point we find

\begin{eqnarray}
h^2 &=&|\mathbf{r}\times \mathbf{u}|^2=r^2u_\varphi
^2|_{perihelion}=r^2(-c^2-u_4^2)|_{perihelion}  \nonumber \\
&=&r^2[-c^2-(\varepsilon -b_0r-b_1\ln r+b_2\frac 1r+b_3\frac
1{2r^2}+...)^2]|_{perihelion} \nonumber\\
\label{r16}
\end{eqnarray}

Since $b_i$ are the coefficients that are independent from distance $r$,
integral constant $h$ and integral constant $\varepsilon $, they take the
same values for various cases which have various man-controlled parameters $h
$ and $\varepsilon $. Now we consider two extreme cases.

\textit{First case: Bob is at rest forever.}

According to the Newton's first law of motion, the interaction between them
must completely vanishes. Since Bob' speed $v$ should not depend on the
distance, according to Eq.(\ref{r15}), a reasonable solution may be $%
r=\infty $, $b_0=0$, $b_1=0$ and $\varepsilon =0$. To note that the values
of $b_0$ and $b_1$ do not depend on this extreme case.

\textit{Second case: with }$h\rightarrow 0$, \textit{Bob passes the
perihelion point about Alice with a distance }$r\rightarrow 0$\textit{.}

According to Eq.(\ref{r16}), a reasonable solution may be that the all
coefficients $b_i$ are zero but except $b_2$.

Therefore we obtain

\begin{eqnarray}
\frac{\mathbf{f}}{u_4} &=&b\frac{\mathbf{r}}{r^3}  \label{r17} \\
u_4 &=&\varepsilon +b\frac 1r  \label{r18}
\end{eqnarray}
where the subscript of $b_2$ has been dropped. Likewise for Alice, we have

\begin{eqnarray}
\frac{\mathbf{f}^{\prime }}{u_4^{\prime }} &=&a\frac{\mathbf{r}}{r^3}
\label{r19} \\
u_4^{\prime } &=&\varepsilon ^{\prime }+a\frac 1r  \label{r20}
\end{eqnarray}
where $a$ and $\varepsilon ^{\prime }$ are coefficients.

Differentiating Eq.(\ref{r7}) with respect to time interval $dt$ gives the
center acceleration as

\begin{eqnarray}
\frac{d\mathbf{u}_c}{dt} &=&0=(m\frac{d\mathbf{u}}{dt}+M\frac{d\mathbf{u}%
^{\prime }}{dt})/(m+M)  \nonumber \\
&=&(\frac{icm}{u_4}\frac{d\mathbf{u}}{d\tau }+\frac{icM}{u_4^{\prime }}\frac{%
d\mathbf{u}^{\prime }}{d\tau ^{\prime }})/(m+M)  \label{r21}
\end{eqnarray}
where we have used $u_4=dx_4/d\tau =icdt/d\tau $ , $u_4^{\prime
}=dx_4^{\prime }/d\tau ^{\prime }=icdt/d\tau ^{\prime }$ . Substituting Eq.(%
\ref{r17}) and Eq.(\ref{r19}) into Eq.(\ref{r21}), we get

\begin{equation}
\frac{icm}{u_4}\mathbf{f}+\frac{icM}{u_4^{\prime }}\mathbf{f}^{\prime }=icm%
\frac{b\mathbf{r}}{r^3}+icM\frac{a\mathbf{r}}{r^3}=0  \label{r22}
\end{equation}
This equation leads to

\begin{equation}
\frac{icb}M=-\frac{ica}m=K  \label{r24}
\end{equation}
where $K$ is a constant. Then Eq.(\ref{r5}) and Eq.(\ref{r6}) may be
rewritten as

\begin{equation}
Bob:\quad m\frac{d\mathbf{u}}{dt}=K\frac{mM\mathbf{r}}{r^3}\qquad \qquad
\label{r25}
\end{equation}

\begin{equation}
Alice:\quad M\frac{d\mathbf{u}^{\prime }}{dt}=-K\frac{mM\mathbf{r}}{r^3}%
\qquad \qquad  \label{r26}
\end{equation}
If $K$ takes a negative constant, then, the above equations show that Bob is
attracted by Alice with Newton's universal gravitation force. \textit{But we
do not want to make this conclusion at once, because there are still a few
problems among them.}

\subsection{Coulomb's force}

\label{Coulf}In this subsection, we study Coulomb's force by based on our
the most simple model: all particles are composed of identical
constituents---Dollons.

From the above subsection, now we can manifestly interpret the quantity $f$
as the 4-vector force exerting on a Dollon of Bob. It is a natural idea to
think of that Dollon has two kinds of charges: positive and negative. If Bob
and Alice are separated by a far distance, and $f$ is the force acting on a
positive Dollon in Bob, then $-f$ is the force acting on a negative Dollon
in Bob. Regardless of the internal forces in Bob, it follows from Eq.(\ref
{r2}) that the motion of the ith Dollon is governed by

\begin{equation}
\frac{du^{(i)}}{d\tau ^{(i)}}=f^{(i)}\quad or\quad \frac{du^{(i)}}{dt}=\frac{%
icf^{(i)}}{u_4^{(i)}}  \label{c1}
\end{equation}
where $d\tau ^{(i)}$, $u^{(i)}$ and $f^{(i)}$ denote the proper time
interval, 4-vector velocity and 4-vector force acting on the ith Dollon,
respectively. Taking sum over all Dollons in Bob, we get

\begin{eqnarray}
\sum\limits_{i=1}^m\frac{du^{(i)}}{dt} &=&\frac
d{dt}[\sum\limits_{i=1}^mu^{(i)}]=\frac{d(mu_c)}{dt}  \label{c2} \\
\sum\limits_{i=1}^m\frac{icf^{(i)}}{u_4^{(i)}} &\simeq &q\frac{icf_c}{u_{c4}}
\label{c3}
\end{eqnarray}
where $u_c$ is the 4-vector velocity of the center of Bob, $u_{c4}$ denotes
its 4th component, $q$ denotes the net charges of Bob, $f_c$ denotes the
4-vector force acting on the Dollon located at the center of Bob ( this
Dollon may be virtual one because it features the average action ).
Combining Eq.(\ref{c2}) and Eq.(\ref{c3}) with Eq.(\ref{c1}), we obtain

\begin{equation}
\frac{d(mu_c)}{dt}=q\frac{icf_c}{u_{c4}}\quad or\quad \frac{d(mu_c)}{d\tau _c%
}=qf_c  \label{c4}
\end{equation}
where we neglect the approximation in Eq.(\ref{c3}).

Like that in the above subsection, the First Law must be valid for the
composite system of Bob and Alice, in other words, when they are separated
from a infinite distance they are isolated, whereas they go to nearest
points they should not touch each other, these requirements lead to

\begin{eqnarray}
Bob &:&\frac{\mathbf{f}_c}{u_{c4}}=b\frac{\mathbf{r}}{r^3}\qquad
u_{c4}=\varepsilon +b\frac 1r  \label{c5} \\
Alice &:&\frac{\mathbf{f}_c^{\prime }}{u_{c4}^{\prime }}=a\frac{\mathbf{r}}{%
r^3}\qquad u_{c4}^{\prime }=\varepsilon ^{\prime }+a\frac 1r  \label{c6}
\end{eqnarray}
where $\varepsilon ,\varepsilon ^{\prime },$ $b$ and $a$ are coefficients.
Without loss of generality, we have

\begin{equation}
\frac{d\mathbf{u}_c}{dt}=0=(\frac{ic}{u_{c4}}\frac{d(m\mathbf{u}_c)}{d\tau _c%
}+\frac{ic}{u_{c4}^{\prime }}\frac{d(M\mathbf{u}_c^{\prime })}{d\tau
_c^{\prime }})/(m+M)  \label{c7}
\end{equation}
Substituting Eq.(\ref{c5}) and Eq.(\ref{c6}) into Eq.(\ref{c7}), we get

\begin{equation}
\frac{ic}{u_{c4}}q\mathbf{f}_c+\frac{ic}{u_{c4}^{\prime }}q^{\prime }\mathbf{%
f}_c^{\prime }=icq\frac{b\mathbf{r}}{r^3}+icq^{\prime }\frac{a\mathbf{r}}{r^3%
}=0  \label{c8}
\end{equation}
where $q^{\prime }$ denotes the net charges of Alice. This equation leads to

\begin{equation}
\frac{icb}{q^{\prime }}=-\frac{ica}q=k  \label{c9}
\end{equation}
where $k$ is a constant. Then the motions of Bob and Alice are governed by

\begin{equation}
Bob:\quad m\frac{d\mathbf{u}}{dt}=k\frac{qq^{\prime }\mathbf{r}}{r^3}\qquad
\qquad  \label{c10}
\end{equation}

\begin{equation}
Alice:\quad M\frac{d\mathbf{u}^{\prime }}{dt}=-k\frac{qq^{\prime }\mathbf{r}%
}{r^3}\qquad \qquad  \label{c11}
\end{equation}
The 4th component equations corresponding to the above equations express the
energy change rates of Bob and Alice, they are not independent components.

The Eq.(\ref{c10}) and Eq.(\ref{c11}) are known as the Coulomb's forces.

\subsection{Gravitational force}

\label{gravf}If Bob and Alice are two atoms with neutral net charges, the
Coulomb's force between them will vanish off. But, precisely, this is not
true, the inspection of Eq.(\ref{c3}) tells us that the net interaction
between them still remains when the atoms are considered as composite
systems.

In this paper, planet, stone, molecule, atom and nucleus are all regarded as
composite systems composed of Dollons, the constituents of the composite
systems move about their centers. If Bob and Alice are two planets with
neutral net charges, then it is reasonable \textit{to assume that the net
force acting on Bob is proportional to the number of Dollons in Bob}, Eq.(%
\ref{c3}) reads

\begin{equation}
\sum\limits_{i=1}^m\frac{icf^{(i)}}{u_4^{(i)}}=g\frac{icmf_c}{u_{c4}}
\label{c12}
\end{equation}
where $g$ is a very very small proportional coefficient. Then the motion of
Bob is given by

\begin{equation}
\frac{d(mu_c)}{dt}=g\frac{icmf_c}{u_{c4}}\quad or\quad \frac{d(mu_c)}{d\tau
_c}=gmf_c  \label{c13}
\end{equation}

In analogy with the above subsections, we may obtain the motion equations of
Bob and Alice, they are governed by

\begin{equation}
Bob:\quad m\frac{d\mathbf{u}}{dt}=-G\frac{mM\mathbf{r}}{r^3}  \label{c14}
\end{equation}

\begin{equation}
Alice:\quad M\frac{d\mathbf{u}^{\prime }}{dt}=G\frac{mM\mathbf{r}}{r^3}
\label{c15}
\end{equation}
where $G$ is a constant proportional to $g$.

The $m$ and $M$ has been identified or defined as the masses by employing
Dollon mass as a unit when we count the Dollon numbers in Bob or Alice. The
Eq.(\ref{c14}) and Eq.(\ref{c15}) are known as the Newton's universal
gravitational forces.

Why is the net force of Bob attractive ? This may be explained as that
electrons with light masses move always around massive nuclei, the
attraction is a little bigger than the repulsion between two atoms separated
by a far distance. In this formulation, the gravitational force possesses
statistic meanings.

\subsection{Nuclear force}

\label{nuclf}We use the most simple model--- all particles are composed of
identical Dollons--- to study nuclear force, to fulfill the conceptual
development boosted by the Newton's first law of motion.

Now consider that Bob and Alice are two nucleons composed of Dollons. If Bob
and Alice go closely in a distance comparable with the sizes of them, then
it is clear that Eq.(\ref{c3}) turns to be inadequate, their polarization
can provide a strong interaction, while the effect of net charges between
their centers becomes to be trivial. The strong interaction is regarded as
the nuclear force in this paper. Therefore, the strong nuclear force is
charge-independent, it only comes into play when the nucleons are very close
together, and it drops rapidly to Coulomb's force for far distance, we know
from experiments that the sensitive distance is about $10^{-15}m$.

As mentioned above, the ith Dollon in Bob is governed by

\begin{equation}
\frac{d\mathbf{u}^{(i)}}{d\tau ^{(i)}}=\mathbf{f}^{(i)}\qquad \qquad \quad 
\frac{du_4^{(i)}}{d\tau ^{(i)}}=-\frac{\mathbf{u}^{(i)}\cdot \mathbf{f}^{(i)}%
}{u_4^{(i)}}  \label{c16}
\end{equation}
Then the motion of Bob is given by

\begin{eqnarray}
\frac{d(m\mathbf{u}_c)}{dt} &=&ic\sum\limits_{i=1}^m\frac{\mathbf{f}^{(i)}}{%
u_4^{(i)}}  \label{c17} \\
\frac{d(mu_{c4})}{dt} &=&-ic\sum\limits_{i=1}^m\frac{\mathbf{u}^{(i)}\cdot 
\mathbf{f}^{(i)}}{[u_4^{(i)}]^2}  \label{c18}
\end{eqnarray}
where

\begin{equation}
mu_c=\sum\limits_{i=1}^mu^{(i)}  \label{c19}
\end{equation}
where $u_c$ can be understood as the velocity of momentum center ( see Eq.(%
\ref{c19} ), but $u_c$ is not the \textit{relativistic} velocity of the
geometrical center of Bob, the \textit{relativistic} velocity of the
geometrical center of Bob is defined by using its geometrical center proper
time, i.e., $u_{center}=dx_{center}/d\tau _{center}$, thus we have to
establish their relation by introducing a correcting factor $\lambda $ so
that $u_c=\lambda u_{center}$, i.e.,

\begin{equation}
\sum\limits_{i=1}^mu^{(i)}=mu_c=\lambda mu_{center}  \label{c20}
\end{equation}

In the following we drop the subscript ''center'' when without confusion,
then above equations can be rewritten as

\begin{eqnarray}
\frac{d(m\lambda \mathbf{u})}{dt} &=&ic\sum\limits_{i=1}^m\frac{\mathbf{f}%
^{(i)}}{u_4^{(i)}}  \label{c21} \\
\frac{d(m\lambda u_4)}{dt} &=&-ic\sum\limits_{i=1}^m\frac{\mathbf{u}%
^{(i)}\cdot \mathbf{f}^{(i)}}{[u_4^{(i)}]^2}  \label{c22}
\end{eqnarray}

To note that the right side of Eq.(\ref{c22}) is the rate at which the
forces do works on Bob, then the quantity $m\lambda u_4$ in the left side
should be ''energy'', thus we can define the energy as

\begin{eqnarray}
E &=&-ic\lambda mu_{c4}=m_r\lambda c^2  \label{c23} \\
m_r &=&\frac m{\sqrt{1-v^2/c^2}}  \label{c24}
\end{eqnarray}
where $u_{c4}=ic/\sqrt{1-v^2/c^2}$, $v$ is the classical speed of the
geometrical center of Bob, $m_r$ is the relativistic mass, while $m$ is the
rest mass. Eq.(\ref{c23}) is known as the energy mass relationship. but Eq.(%
\ref{c23}) has a little difference from Einstein's mass-energy relationship.
Our energy formula contains a factor $\lambda $ that represents the internal
motion of Dollons in Bob, obviously, $\lambda \geq 1$, this can be seen
clearly from Eq.(\ref{c20}), in other words, even if the center is at rest,
the internal constituents can still have relativistic energies.

In dealing with nuclear reaction, in many textbooks, mass defect is
understood as the decrease in total relativistic mass, even if all nuclei
seem to be at rest before or after the nuclear reaction--- the total
relativistic masses should not have apparent change. We have been puzzled by
these statements for a long time. Now the reasons are clear, no relativistic
masses change but $\lambda $ changes in these cases, in other words, the
internal energy of particle has changed. $\lambda $ is a physical quantity
sensitive to the internal structure of a particle, is a criteria for
particle being elementary or not.

Consider that a hadron possesses net charge $q$, we can naturally image that
the charge distributes in several parts inside the hadron, assuming three
parts, the three parts have net charges denoted by $I_q$, $B_q$, and $S_q$
respectively, then

\begin{equation}
q=I_q+B_q+S_q  \label{q1}
\end{equation}
Comparing with the Gell-Mann-Nishijima relation

\begin{equation}
q=I_3+\frac{B+S}2  \label{q2}
\end{equation}
we can understand the conservations of isospin $I_3$, baryon number $B$ and
strangeness number $S$ with four remarks: (1) the three parts inside the
hadron are insulated from one another, no charge transports from one to
another. (2) during collision of hadrons, only identical parts impact or
touch each other, with exchanging net charges. (3) the mass of the hadron
seems to depend primarily on the masses of the parts inside the hadron,
weakly on the net charges of the parts. (4) if we assign the quantum states
of quarks $u$, $s$ and $d$ to the three parts, the quark model seems to be
improved in a manner that we can avoid the fractional charges of the quarks.

\subsection{Determining the 4-vector $X$}

\label{X} In the preceding subsections, we have mentioned that the
interaction between Bob and Alice we studied happens at such instant that
their relative position in the Minkowski's space is denoted by a 4-vector $%
X=(\mathbf{r},X_4)=(\mathbf{r},ic\triangle t)$, $X$ satisfies the orthogonal
relation simultaneously

\begin{equation}
u\cdot X=0\qquad u^{\prime }\cdot X=0  \label{c30}
\end{equation}
The purpose of choosing this instant is to meet the convenience that $X$
parallels to $f$ and $f^{\prime }$ simultaneously, because

\begin{eqnarray}
Bob &:&\qquad u\cdot f=u_\mu f_\mu =0  \label{c31} \\
Alice &:&\qquad u^{\prime }\cdot f^{\prime }=u_\mu ^{\prime }f_\mu ^{\prime
}=0  \label{c32}
\end{eqnarray}
See Eq.(\ref{r8})-(\ref{r11}). Eq.(\ref{c30}) can be rewritten in the form
of inner product of two vectors as

\begin{equation}
|u|\cdot |X|\cosh (u,X)=|u^{\prime }|\cdot |X|\cosh (u^{\prime },X)=0
\label{c33}
\end{equation}
This leads to two solutions given by

\begin{eqnarray}
|X| &=&\sqrt{r^2-(c\triangle t)^2}=0  \label{c34} \\
\cosh (u,X) &=&\cosh (u^{\prime },X)=0  \label{c35}
\end{eqnarray}
Eq.(\ref{c33}) again leads to two solutions given by

\begin{equation}
r=c\triangle t\qquad r=-c\triangle t  \label{c36}
\end{equation}
The first solution expresses that the force acting on Bob is retarded by
time $\triangle t=r/c$, the second one expresses that the action is
preceded. Our choice is the first one which gives an effect that follows the
cause.\ We know, this retarded time is just the time needed for the
propagation of interaction from Alice to Bob, the propagation speed is $c$,
no mater what kind of interaction.

Eq.(\ref{c35}) represents the orthogonal relationship.

Therefore, the interaction happens at such instant that either in retarded
state or in orthogonal state, or mixture.

\section{Minkowski's space}

In preceding sections, we have realized that relativistic Newton's second
law and forces can be derived from Newton's first law and the magnitude
formula of 4-vector velocity of particle. The formula is given by

\begin{equation}
u_\mu u_\mu =-c^2  \label{m0}
\end{equation}
in a Minkowski's space. It is noted that all particles satisfy the above
equation, it then is regarded as the origin of the particle invariance. We
wonder at what is the essence of the Minkowski's space. In this section we
shall discuss the Minkowski's space, for this purpose we need to establish a
standard method for describing the motion of particle in space-time. Our
construction follows four steps.

\subsection{First Step: we are lazy}

Suppose Alice is a pretty girl being famous for her fast running records, we
state some her records here in a story ( in imagination ).

(1) Jan., 1, 2001, 10:00 am, sportsground in BUAA, Beijing. In a time
interval $\Delta t=10s$ Alice ran a straight line distance $\Delta l_1=100m$
at a constant speed $v_1=10m/s$.

This data can be given in physical terms by

\begin{equation}
\Delta l_1=v_1\Delta t  \label{m1}
\end{equation}
It can be rewritten either as

\begin{equation}
\Delta x_1^2+\Delta y_1^2=(v_1\Delta t)^2  \label{m2}
\end{equation}
or as

\begin{equation}
\Delta x_1^2+\Delta y_1^2-(v_1\Delta t)^2=0  \label{m3}
\end{equation}
where $x$ and $y$ denote the coordinate system fixed at the sportsground. By
defining a imaginary quantity $w_1=iv_1t$, the data is given by

\begin{equation}
\Delta x_1^2+\Delta y_1^2+\Delta w_1^2=0  \label{m4}
\end{equation}
We appreciate the simplicity and beauty of its form.

It is also our favorite manner to mark the running process in a graph with
three mutually perpendicular axes $x,y$ and $w=iv_1t$. The distance from the
starting point to the final point in this coordinate system equals to zero
because of Eq.(\ref{m4}). This graph we called as ''20010101 Graph''.

(2) Jan., 2, 2001, 10:00 am, sportsground in BUAA, Beijing. In a time
interval $\Delta t=10s$ Alice ran a straight line distance $\Delta l_2=90m$
at a constant speed $v_2=9m/s$.

This data is given in physical terms by

\begin{equation}
\Delta x_2^2+\Delta y_2^2+\Delta w_2^2=0\qquad \Delta w_2=iv_2\Delta t
\label{m5}
\end{equation}

We directly mark this day running process in the yesterday's 20010101 Graph,
we are lazy to draw a new graph.

(3) Jan., 3, 2001, 10:00 am, sportsground at BUAA, Beijing. In a time
interval $\Delta t=10s$ Alice ran a straight line distance $\Delta l_3=95m$
at a constant speed $v_3=9.5m/s$.

We also directly mark the running process in the 20010101 Graph.

Bob was also a good runner, in a time interval $\Delta t=10s$ Bob ran a
straight line distance $\Delta l_b=105m$ at a constant speed $v_b=10.5m/s$.

We also directly mark the running process in the 20010101 Graph.

In fact, their running records all are marked in the 20010101 Graph.

\subsection{Second Step: Establishing a temporary standard frame}

Because of laziness, we only use the 20010101 Graph to record the running
data, it has actually become a temporary standard frame, all motions can be
marked or calculated in the Graph, it is much connivent for describing any
movement.

To note that $w=iv_1t$, the $w$ axis in the Graph has involved the speed $%
v_1 $ created by Alice on Jan 1, 2001. Thus we find that the geometrical
distance $\Delta s_2$ from the starting point to the final point in the
Graph on the second day for Alice is not equal to zero.

\begin{equation}
\Delta s_2^2=\Delta x_2^2+\Delta y_2^2+\Delta w_1^2\neq 0\qquad \Delta
w_1=iv_1\Delta t  \label{m6}
\end{equation}
It is clear after comparing with Eq.(\ref{m5}). So do for Bob, the distance $%
\Delta s_b$ for Bob in the 20010101 Graph is given by

\begin{eqnarray}
\Delta s_b^2 &=&\Delta x_b^2+\Delta y_b^2+\Delta w_1^2  \nonumber \\
&=&\Delta x_b^2+\Delta y_b^2-v_1^2\Delta t^2  \nonumber \\
&=&\Delta x_b^2+\Delta y_b^2-v_b^2\Delta t^2+v_b^2\Delta t^2-v_1^2\Delta t^2
\nonumber \\
&=&v_b^2\Delta t^2-v_1^2\Delta t^2  \nonumber \\
&=&-v_1^2\Delta t^2(1-v_b^2/v_1^2)  \label{m7}
\end{eqnarray}

\begin{equation}
\Delta x_b^2+\Delta y_b^2-v_1^2\Delta t^2=-v_1^2\Delta t^2(1-v_b^2/v_1^2)
\label{m8}
\end{equation}
Dividing the two sides of the above equation by $\Delta t^2(1-v_b^2/v_1^2)$,
we get

\begin{eqnarray}
-v_1^2 &=&\left( \frac{\Delta x_b/\Delta t}{\sqrt{1-v_b^2/v_1^2}}\right)
^2+\left( \frac{\Delta y_b/\Delta t}{\sqrt{1-v_b^2/v_1^2}}\right) ^2 
\nonumber \\
&&+\left( \frac{iv_1}{\sqrt{1-v_b^2/v_1^2}}\right) ^2  \label{m9}
\end{eqnarray}
Defining modified velocity

\begin{eqnarray}
u_x &=&\frac{v_x}{\sqrt{1-v_{}^2/v_1^2}}  \label{m10} \\
u_y &=&\frac{v_y}{\sqrt{1-v_{}^2/v_1^2}}  \label{m11} \\
u_w &=&\frac{iv_1}{\sqrt{1-v_{}^2/v_1^2}}  \label{m12}
\end{eqnarray}
where $v^2=v_x^2+v_y^2$, we have dropped the subscript $b$ that indicates
Bob, then Eq.(\ref{m9}) is given by

\begin{equation}
u_x^2+u_y^2+u_w^2=-v_1^2  \label{m13}
\end{equation}
The modified velocity of Bob in the 20010101Graph is based on the Alice's
best speed record $v_1$. In fact, Any one, any body or any particle, their
modified velocity in the 20010101 Graph satisfies Eq.(\ref{m13}).

\subsection{Third Step: Standard Graph based on the light speed}

Because of convenience, it has become a habit for us to use the 20010101
Graph to mark the all motions of any body. It is sure that not all
scientists in the world like Alice, then we gradually recognize that we need
a permanent runner for establishing a standard graph. Now we had to face a
new task: to look for a new hero.

It was said that the light, an element of the nature, is the fastest runner,
whenever and whereever its speed is $3\times 10^8m/s$. We do not hesitate to
use the light speed to replace Alice's speed, and setup a new frame called
''Standard Graph'', the Standard Graph contains four mutually perpendicular
axes $x$, $y$, $z$ and $w=ict$ ( we can draw several partial frames to
assemble the whole frame ). From then, any motions can be described in the
Standard Graph with the space-time $(x,y,z,ict)$ or $(x_1,x_2,x_3,x_4=ict)$.
In analogy with Eq.(\ref{m10})-(\ref{m13}), defining modified velocity

\begin{eqnarray}
u_x &=&\frac{v_x}{\sqrt{1-v_{}^2/c^2}}\qquad u_y=\frac{v_y}{\sqrt{%
1-v_{}^2/c^2}}  \label{m14} \\
u_z &=&\frac{v_z}{\sqrt{1-v_{}^2/c^2}}\qquad u_w=\frac{ic}{\sqrt{1-v_{}^2/c^2%
}}  \label{m15}
\end{eqnarray}
where $v^2=v_x^2+v_y^2+v_z^2$, we obtain

\begin{equation}
u_x^2+u_y^2+u_z^2+u_w^2=-c^2\qquad u_\mu u_\mu =-c^2  \label{m16}
\end{equation}
The Standard Graph is just the Minkowski's space, the 4-vector velocity $%
u=\{u_\mu \}$ is known as the relativistic velocity.

\subsection{Fourth Step: Transformations}

We immediately recognize that physics holds its validity only in the
Standard Graph (involving with the light speed), rather than in the 20010101
Graph (involving with the Alice's speed), this situation can be explained by
the fact that all physical quantities and their measurements are defined on
facilities whose principles are based on the light directly or indirectly,
for example, the ''meter'' and ''second'' are defined on the light speed
directly.

If we do not hope that one graph has advantage over than another, then the
transformation between the Standard Graph and 20010101 Graph will be given by

\begin{equation}
x=x_1,y=y_1,z=z_1,t=\frac{v_1}ct_1  \label{m16a}
\end{equation}
where the subscript 1 denotes in the 20010101 Graph. It means we need to
redefine all physical quantities such as rod and clock in the 20010101
Graph, do not use the light.

\section{Dynamics in the Hilbert space}

A complete inner product space is called a Hilbert space. Our experience in
the preceding sections tells us that it is an easy thing to put dynamics
into the Hilbert space if we have an invariant quantity. The formalism of
the interaction can be derived from some basic laws, it is strongly based on
concrete instances.

\section{Discussion}

In the section \ref{basicf}, the Newton's first law of motion means that the
4-vector average velocity of an isolated system remains at rest or in
motion. This explanation is based on the definition of average velocity
given by

\begin{equation}
u_c=\frac 1S\sum\limits_i^Su^{(i)}  \label{d1}
\end{equation}
where $S$ denotes the number of Dollons in the system as in Eq.(\ref{r7}).
The Newton's first law of motion becomes a sort of strong constraint,
inevitably leads to action reaction law or momentum conversation law being
valid inside the system, for example, for a rest system composed of two
Dollons Alice and Bob we have

\begin{equation}
\mathbf{u}_a+\mathbf{u}_b=\frac{d\mathbf{x}_a}{d\tau _a}+\frac{d\mathbf{x}_b%
}{d\tau _b}=0\qquad u_{4a}+u_{4b}=const  \label{d2}
\end{equation}
The above equation means that the action and reaction are equal in magnitude
and reverse in directions on the line joining the two particle. But we
immediately wonder at that Alice and Bob have to adjust their proper times $%
d\tau _a$ and $d\tau _b$ from time to time to meet the requirement of Eq.(%
\ref{d2}). That is why we say the First Law is a strong constraint for the
system.

Obviously, the geometrical center of a system is defined by

\begin{equation}
x_{center}=\frac 1S\sum\limits_i^Sx^{(i)}  \label{d3}
\end{equation}
The relativistic 4-vector velocity of the geometrical center of the system
is given by

\begin{equation}
u_{center}=\frac{dx_{center}}{d\tau _{center}}  \label{d4}
\end{equation}
As mentioned in the section \ref{nuclf}, $u_{center}\neq u_c$. Immediately,
we find the Newton's first law of motion can be newly explained by based on
the relativistic 4-vector velocity of the geometrical center of the system,
i.e., the Newton's first law of motion means that the 4-vector velocity of
the geometrical center of an isolated system remains at rest or in motion.
This new explanation implies that the action reaction law for the
relativistic 4-vector forces inside the system are not held [comparing to
Eq.(\ref{d2}) ], but the following expansions for Alice and Bob become
possible.

\begin{eqnarray}
Bob &:&\qquad f_b=AX+Bu_a  \label{d5} \\
Alice &:&\qquad f_a=CX+Du_b  \label{d6}
\end{eqnarray}
where $A$, $B$, $C$ and $D$ are unknown coefficients. All conclusions we
obtained in the preceding sections can be retained or modified by retracing
the route of the paper, in accordance with the section \ref{direction}. The
new explanation seems to be much reasonable, but it is worth further
studying the action reaction law and momentum conversation law which are
confronting with serious troubles, they need special treatment like that for
Ampere's force in electromagnetism.

Another topic we would like to discuss briefly is SU(n) group. Each
infinitesimal transformation of the SU(n) group can be written in the form

\begin{equation}
U=1+if_kH_k  \label{d7}
\end{equation}
As usual, repeated indices must be summed over. Where the real parameter $%
f_k $ are treated as small quantities, $U$ and $H_k$ are matrices which
satisfy the definition of the group

\begin{equation}
UU^{+}=(1+if_kH_k)(1-if_kH_k^{+})=1  \label{d8}
\end{equation}
We recall from Eq.(\ref{20}) that Dirac equation was derived from the
following equation

\begin{equation}
\lbrack a_{\nu lj}P_\nu \psi ^{(l)}+i\delta _{lj}mc\psi ^{(l)}][a_{\mu
jk}P_\mu \psi ^{(k)}-i\delta _{jk}mc\psi ^{(k)}]=0  \label{d9}
\end{equation}
It is much impressive that Eq.(\ref{d8}) and Eq.(\ref{d9}) have a similar
form, especially when we let a matrix $E$ to absorb the right side of Eq.(%
\ref{d8}), i.e.

\begin{equation}
(1+iE+if_kH_k)(1-iE^{+}-if_kH_k^{+})=0  \label{d10}
\end{equation}
From this comparison we may understand why the SU(n) group could embed in
quantum mechanics in a obscure way. This situation arouses our interest to
measure a new group whose matrices satisfy

\begin{equation}
Z_j=1+if_kY_{jk}\qquad Z_jZ_j^{+}=0  \label{d11}
\end{equation}
We believe this new group has even more direct relations with quantum
mechanics.

\section{Conclusions}

It is important to recognize that physics must be invariant for composite
particles and their constituent particles, only one physical formalism
exists for any particle, this requirement is called particle invariance.

Under the particle invariance, it is rather remarkable to find that
Klein-Gordon equation and Dirac equation can be derived from the
relativistic Newton's second law of motion on different conditions
respectively, thus only one formalism is necessary for particle, the
relativistic Newton's second law is regarded as one which suitable for any
kinds of particles.

We point out that the Coulomb's force and gravitational force on a particle
always act in the direction orthogonal to the 4-vector velocity of the
particle in 4-dimensional space-time, rather than along the line joining a
couple of particles. This inference is obviously supported from the fact
that the magnitude of the 4-vector velocity is kept constant. Maxwell's
equations can be derived from classical Coulomb's force and the magnitude
formula of 4-vector velocity of particle.

Our speculation on the quarks model leads to introduce a new elementary
particle called Dollon to assemble particles such as baryons, mesons and
other composite particles. Instead of quark model, the Dollon model is
better in organizing known data, specially in modelling interactions. It is
found that relativistic Newton's second law and various interactions can be
derived from the Newton's first law of motion and the magnitude formula of
4-vector velocity of particle.

The structure of Minkowski's space is discussed in detail, it indicates that
the magnitude formula of 4-vector velocity of particle is only a geometrical
distance formula (Pythagoras's theorem), so that it is completely free from
any particle property. Any dynamics or dynamical characteristics originated
from the magnitude formula of 4-vector velocity of particle will completely
preserve the particle invariance, i.e., the dynamics do not distinguish
particle species. Thus the magnitude formula of 4-vector velocity of
particle is regarded as the origin of the particle invariance.

Dynamics in Hilbert space can be established in the same way.


\begin{thebibliography}{99}
\bibitem{Harris}  E. G. Harris, Introduction to Modern Theoretical Physics,
Vol.1\&2, (John Wiley \& Sons, USA, 1975).

\bibitem{Zeng}  J. Y. Zeng, Quantum Mechanics, (Science, China, 1981).

\bibitem{Schiff}  L. I. Schiff, Quantum Mechanics, Third edition,
(McGraw-Hill, USA, 1968).

\bibitem{Kaplan}  D. Kaplan and L. Glass, Understanding Nonlinear Dynamics,
(Springer-Verlag, NY, 1995).

\bibitem{Nachmann}  O. Nachmann, Elementary Particle Physics,
(Springer-Verlag, Berlin, 1990).

\bibitem{Halzen}  F. Halzen and A. Martin, Quarks and Leptons: An
Introductory Course in Modern Particle Physics, (John Wiely \& Sons, NY,
1984).

\bibitem{Cui89}  H. Y. Cui, College Physics (in Chinese), \textbf{4},
13(1989).

\bibitem{Cui92}  H. Y. Cui, College Physics (in Chinese), \textbf{10},
31(1992).

\bibitem{Cui01114}  H. Y. Cui, eprint, quant-ph/0102114, (2001).

\bibitem{Cui0172}  H. Y. Cui, eprint, quant-ph/0108072, (2001).

\bibitem{Cui0173}  H. Y. Cui, eprint, physics/0102073, (2001).

\bibitem{Cui0123}  H. Y. Cui, eprint, physics/0103023, (2001).

\bibitem{Cui0119}  H. Y. Cui, eprint, physics/0103019, (2001).

\bibitem{Lamb}  W. E. Lamb, Jr., Am. J. Phys., 69, 4,(2001).
\end{thebibliography}
\end{document}